\documentclass[fleqn,usenatbib]{mnras}

\usepackage{newtxtext,newtxmath}

\usepackage[T1]{fontenc}

\DeclareRobustCommand{\VAN}[3]{#2}
\let\VANthebibliography\thebibliography
\def\thebibliography{\DeclareRobustCommand{\VAN}[3]{##3}\VANthebibliography}

\usepackage{graphicx}   	
\usepackage{amsmath}	    
\usepackage{gensymb}	    
\usepackage{anyfontsize}    
\usepackage{xcolor}         

\newcommand{\thisstar}{HD~218261}
\newcommand{\thisstarA}{HD~218261~A}
\newcommand{\thisstarB}{HD~218261~B}

\newcommand\masyr{\ensuremath{\text{mas}\,\text{yr}^{-1}}}
\newcommand\cms{\ensuremath{\text{cm}\,\text{s}^{-2}}}

\definecolor{my_color}{HTML}{CF0000}
\newcommand\bmaroon{}

\title[Serendipitous MIRI white dwarf observation]{Serendipitous observation of a white dwarf companion to a JWST/MIRI coronagraphic calibrator}

\author[Venner et al.]{Alexander Venner,$^{1}$\thanks{E-mail: alexandervenner@gmail.com}
Mary Anne Limbach,$^{2}$
Mathilde M\^alin,$^{3,4,5}$
Simon Blouin,$^{6}$
Anthony Boccaletti,$^{5}$
\newauthor{Logan A. Pearce$^{2}$}
\\
$^{1}$Centre for Astrophysics, University of Southern Queensland, Toowoomba, QLD 4350, Australia\\
$^{2}$Department of Astronomy, University of Michigan, Ann Arbor, MI 48109, USA\\
$^{3}$Space Telescope Science Institute, Baltimore, MD 21218, USA \\
$^{4}$Department of Physics \& Astronomy, Johns Hopkins University, Baltimore, MD 21218, USA\\
$^{5}$LESIA, Observatoire de Paris, Université PSL, CNRS, Sorbonne Université, Univ. Paris Diderot, Sorbonne Paris Cité, 92195 Meudon, France \\
$^{6}$Department of Physics and Astronomy, University of Victoria, Victoria, BC V8W 2Y2, Canada \\
} 

\date{Accepted 2024 November 14. Received 2024 November 14; in original form 2024 October 15}

\pubyear{2024}

\begin{document}
\label{firstpage}
\pagerange{\pageref{firstpage}--\pageref{lastpage}}
\maketitle

\begin{abstract}

We present the unplanned detection of a white dwarf companion to the star HD~218261 in mid-infrared (10-16~$\mu$m) observations with JWST/MIRI. This star was observed as a calibrator for coronagraphic observations of the exoplanet host HR~8799. HD~218261~B has only previously been detected by \textit{Gaia}, and only in visible light. We confidently detect the companion in the mid-infrared, where it is less luminous than the primary by a factor of $\sim$10$^4$. The visible and mid-infrared photometry are consistent with a white dwarf of $T_\text{eff}\approx10000$~K, $M\approx0.8~M_\odot$, though observation of its optical spectrum is required to precisely constrain its physical parameters. These results demonstrate that precise mid-infrared photometry of white dwarf companions to bright stars can be obtained with MIRI, opening up new possibilities for studying white dwarfs in close binaries.

\end{abstract}




\section{Introduction} \label{sec:intro}

The recently launched James Webb Space Telescope \citep[JWST;][]{JWST} has already revolutionised infrared (IR) astrophysics. A vast array of results have been achieved through its first years of observations. Furthermore, the unprecedented sensitivity of JWST across infrared wavelengths means that there is also significant potential for serendipitous \bmaroon{science results from} observations intended for other purposes. A notable example of this phenomenon is the discovery of some of the faintest and most distant brown dwarfs ever detected in deep-field observations intended for extragalactic astrophysics \citep[ex.][]{Nonino2023, Langeroodi2023, Hainline2024}.

The Mid-InfraRed Instrument (MIRI) on JWST is the most sensitive astronomical instrument operating in mid-IR wavelengths, and has been designed to break new ground in various areas of astrophysical research \citep{Rieke2015}. MIRI is equipped four coronagraphs (three 4QPM coronagraphs between 10-16~$\mu$m\textbf{}, and a 23~$\mu$m Lyot mask), giving it an unparalleled capability for high-contrast imaging in the mid-IR \citep{Boccaletti2015, Boccaletti2022}.

In this Letter we report the serendipitous observation of a white dwarf companion to the star \thisstar{} in JWST/MIRI coronagraphic data, collected as part of the observations presented in \citet{Boccaletti2024}. We highlight how this demonstrates the viability of mid-infrared observations of white dwarfs in high-contrast binaries with JWST.

\section{Observations} \label{sec:observations}

\subsection[Gaia]{\textit{Gaia}}

\thisstar{} (HR~8792, HIP~114096) is a $V=6.4$ F8.5V star at a distance of $29.31\pm0.02$~pc \citep{Gray2001, GaiaDR3}. The star was not known to be in any way unusual until a comoving stellar companion was discovered by \textit{Gaia} \citep{Gaia} at a projected separation of 5.24~arcseconds \citep[$\sim$150~AU;][]{ElBadry2021}. In Table~\ref{tab:parameters} we summarise the properties of the two stars reported in \textit{Gaia}~DR3 \citep{GaiaDR3}.

\begin{table}
\caption{Properties of \thisstar{}~AB from \textit{Gaia}~DR3.}
\begin{tabular}{lrr}
\hline
Parameter & \thisstarA{} & \thisstarB{} \\
\hline
Source ID & 2831490694929214464 & 2831490694928280576 \\
R.A. $\alpha$ & 23:06:31.89 & 23:06:31.53 \\
Declination $\delta$ & +19:54:39.07 & +19:54:37.95 \\
Parallax (mas) & $34.077\pm0.019$ & $33.98\pm0.10$ \\
$\mu_\alpha$ (\masyr{}) & $+286.890\pm0.023$ & $+281.90\pm0.15$ \\
$\mu_\delta$ (\masyr{}) & $+5.057\pm0.019$ & $-1.41\pm0.10$ \\
\textit{G} (mag) & $6.314\pm0.0028$ & $15.026\pm0.0037$ \\
\textit{BP} (mag) & $6.582\pm0.0028$ & $14.453\pm0.048$ \\
\textit{RP} (mag) & $5.882\pm0.0038$ & $13.924\pm0.037$ \\
\hline
\end{tabular}
\label{tab:parameters}
\end{table}

The companion star, \thisstarB{}, has an estimated absolute magnitude $M_G=12.69$~mag and \textit{BP-RP} colour $\approx$0.53, properties consistent with a white dwarf \citep[as previously identified by][]{Golovin2024}. However, this star does not appear in existing \textit{Gaia} white dwarf catalogues \citep{GentileFusillo2019, GentileFusillo2021}. The star is brighter in both \textit{BP/RP} compared to the \textit{G} band, which is incongruous since these photometric passbands are approximately complementary \citep{GaiaEDR3.photometry}. Pursuing this, there is evidence that the \textit{Gaia} \textit{BP/RP} photometry for this star is of poor quality; the \textit{Gaia}~DR3 \texttt{phot\_bp\_rp\_excess\_factor} is $2.39$, and following \citet[][equation~6]{GaiaEDR3.photometry} this results in a normalised flux excess factor $C^*=1.21$, which is significantly exceeds $C^*=0$ expected for well-behaved sources. This also fails the $C^*<0.6$ quality cut employed by \citet{GentileFusillo2021}, which explains the absence from their white dwarf catalogue.

We interpret the high value of $C^*$ for \thisstarB{} as evidence for flux contamination from the much brighter primary. The \textit{BP} and \textit{RP} photometry is extracted from a 3.5$\times$2.1~arcsec$^2$ window \citep[][section~9.3]{GaiaEDR3.photometry}, which is comparable to the angular separation of the binary; considering that \thisstarA{} is brighter than B by a factor of $3\times10^4$ in \textit{G}, even a small fraction of stray light could easily swamp the \textit{BP/RP} photometry of its companion. We conclude that the \textit{BP} and \textit{RP} photometry for \thisstarB{} is unreliable.

\subsection{JWST}

\thisstar{} was \bmaroon{selected for observation} with MIRI coronagraphic imaging as part of JWST GTO programme 1194 (PI: Beichman), as a proximate reference star used to calibrate the point spread function (PSF) of the exoplanet host HR~8799 \citep[$\sim$1.24$\degree$ separation,][]{Boccaletti2024}. \bmaroon{\thisstar{} was originally selected for this purpose prior to the 2018 release of \textit{Gaia}~DR2, so the existence of \thisstarB{} was not known when the observation plan was first prepared.}

\bmaroon{\thisstar{} was observed by JWST on 2022-11-08. Images} were acquired with the F1065C, F1140C, and F1550C coronagraphic filters. At a projected separation of 5.2", \thisstarB{} is sufficiently faint and widely separated from \thisstarA{} to be spatially resolved with MIRI and does not affect its use as a reference star for removing the stellar diffraction pattern in observations of HR~8799\bmaroon{, which are mainly found below $<$1". Indeed, \thisstarB{} is fainter than several background sources in the MIRI fields.} However, these observations do allow for serendipitous detection of \thisstarB{} in the mid-IR.

\begin{figure*}
    \centering
    \includegraphics[width=\textwidth]{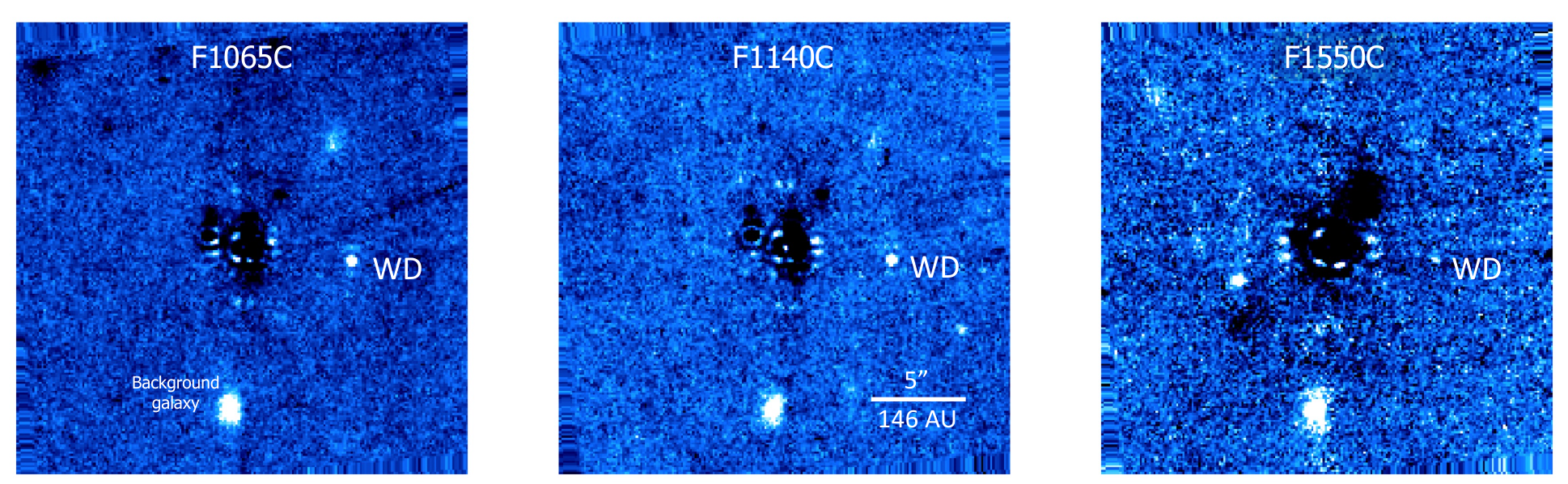}
    \caption{Reference star subtracted images of \thisstar{} in the three MIRI coronagraphic filters (left to right: F1065C, F1140C, F1550C). The full MIRI 4QFM coronagraphic field of view (24” × 24”) is displayed. The white dwarf companion \thisstarB{} is labeled as ``WD''. Negative sources from the HR~8799 field are apparent at small separations.}
    \label{fig:MIRI}
\end{figure*}

As a first step, we take the inverse of the PSF-subtracted images of HR~8799 to inspect the detected sources in the \thisstar{} field. We show the resulting images in Figure~\ref{fig:MIRI}. We clearly observe a source at the companion position expected from \textit{Gaia} ($\rho=$~5.24", PA = 256$\degree$) in the F1065C and F1140C filters, and also weakly detect the same source in F1550C. We also note that the corresponding negative source is quite visible in \citet[][figure 1.1, 1.2]{Boccaletti2024}. This suffices to demonstrate that \thisstarB{} has been detected in the MIRI observations.

Next, we performed a more involved analysis of the MIRI data to extract precise photometry for \thisstarB{}. We started from the calibrated images used in \cite{Boccaletti2024}. We performed a PSF subtraction to eliminate the stellar diffraction pattern of \thisstarA{}, using the commissioning data from JWST~GTO~1037 \citep{Boccaletti2022} for the reference star and applying applied a PCA-based method as in \citet{Boccaletti2024}.
The inner region of the image still shows residual starlight because the reference star was not optimised for the observations of \thisstar{}. However, the processing performs well enough for larger separations.

Following the method described in \citet{Malin2024}, we modelled the PSF of \thisstarB{} with {\sc WebbPSF} \citep{Perrin2014}. We simulate a PSF for each filter taking into account the combination of filter, coronagraphic mask and pupil mask, as well as the position on the detector. We minimise the residuals between the simulated PSF and the data in a 2$\lambda$/D region centered on the object. Minimization is performed using the Nelder-Mead algorithm \citep{Gao2012} with the \texttt{optimize.minimize} function from {\sc scipy} \citep{Virtanen2020}. \bmaroon{To initialise the minimiser}, we chose the position of the brightest pixel and an arbitrary flux of 1 DN/s. A possible $\pm$1-pixel offset is allowed in the minimization process to account for sub-pixel offsets between the model and the simulated PSF.

The coronagraphic transmission is $\approx$1 at the separation of \thisstarB{}, so this has a negligible impact on the measured flux. As a result, we measured the photometry by simply calculating the flux of the best-fit PSF model for each filter. \thisstarA{} was observed using the 9-point dither pattern (as the observing strategy was designed to catch the PSF diversity for the HR~8799 observations). We therefore repeated the PSF model at all dither positions to obtain the uncertainty on the flux measurement.


\section{Results} \label{sec:results}

\begin{table}
\caption{Photometry of \thisstarB{}.}
\begin{tabular}{lll}
\hline
Band / Filter & $\lambda$ ($\mu$m) & Flux ($\mu$Jy) \\
\hline
\textit{BP} & $0.52$ & $5880\pm260$ $^{(a)}$ \\
\textit{G}  & $0.64$ & $3150\pm10$  \\
\textit{RP} & $0.78$ & $6880\pm240$ $^{(a)}$ \\
F1065C & $10.58$ & $40.0\pm3.2$ \\
F1140C & $11.3$ & $26.3\pm2.3$ \\
F1550C & $15.5$ & $17.7\pm2.0$ \\
\hline
\multicolumn{3}{l}{$^{(a)}$ Contaminated by \thisstarA{}.} \\
\end{tabular}
\label{tab:photometry}
\end{table}

In Table~\ref{tab:photometry} we report our extracted MIRI photometry for \thisstarB{} along with existing measurements from \textit{Gaia}~DR3. In the three MIRI coronagraphic filters we retrieve fluxes between \bmaroon{$18-40$}~$\mu$Jy, two orders of magnitude lower than observed in the optical. Qualitatively this can only be explained by a small and hot star, confirming the white dwarf identification of \citet{Golovin2024} based on the poor quality \textit{Gaia} photometry. The mid-IR flux of \thisstarA{} has been measured at 11.6~$\mu$m by WISE \citep{Wright2010} as $0.250\pm0.003$~Jy. Hence the flux contrast with the companion is about $10^{4}$ ($\sim$10~mag) at these wavelengths. The high significance of this detection exemplifies the exceptional imaging capabilities of JWST and its potential for serendipitous science.

Our constraints on the spectral energy distribution of \thisstarB{} are highly incomplete, being that we have only a limited set of visible and mid-IR photometry. Nonetheless, we attempt to use white dwarf spectral models to make a preliminary exploration of its properties. We used the models of \citet{Blouin2018.I, Blouin2018.II} to estimate the atmospheric parameters of \thisstarB{}. We make the assumption that \thisstarB{} has a hydrogen-dominated (DA) spectrum, as this is the most common atmospheric composition among white dwarfs; however, this cannot be proven or falsified with the existing data. We adjusted the effective temperature to match the observed slope between the \textit{Gaia} $G$ band and MIRI F1065C, F1140C, and F1550C filters. We then scaled the model flux to match the observed fluxes, given the distance implied by the \textit{Gaia} parallax, to determine the stellar radius. The surface gravity was self-consistently adjusted using the evolutionary models of \cite{Bedard2020} \bmaroon{to relate the mass and radius of the white dwarf.}

\begin{figure}
    \centering
    \includegraphics[width=\columnwidth]{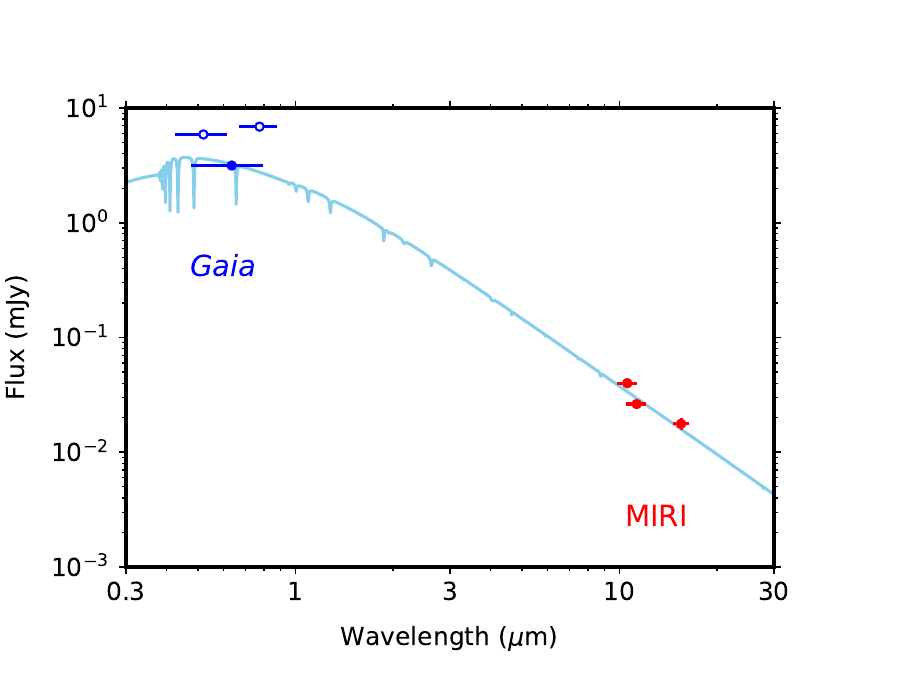}
    \caption{Spectral energy distribution of \thisstarB{}. A model spectrum with $T_\text{eff}=10000$~K, $\log~g=8.3$ adequately reproduces the measured photometry, except for the \textit{Gaia BP/RP} bands (unfilled points).}
    \label{fig:SED}
\end{figure}

\bmaroon{We began iterating through possible white dwarf parameters using} $\log~g=8.0$ (i.e. $10^8$~\cms{}), equivalent to a $M=0.6~M_\odot$ white dwarf lying close to the peak of the mass distribution \citep{Tremblay2016}. However, we found that the resulting model spectra overpredict the mid-infrared flux. Exploring \bmaroon{other} models with lower luminosities, we find that a $T_\text{eff}=10000$~K, $\log~g=8.3$ ($M=0.8~M_\odot$) model spectrum provides a satisfactory fit to the photometry (Figure~\ref{fig:SED}). No model can reproduce the large difference between the \textit{G} and \textit{BP}/\textit{RP} fluxes, consistent with our interpretation that this these measurements are contaminated by the primary.

\bmaroon{As a check for consistency, we can compare the implied age of \thisstarB{} with that of A as both stars are expected to be coeval. Using the \citet{Bedard2020} models, a $T_\text{eff}=10000$~K, $\log~g=8.3$ white dwarf entails a cooling age of $\approx$1.0~Gyr. The main sequence progenitor of a 0.8~$M_\odot$ white dwarf has a mass of $\sim$3.5~$M_\odot$, implying an additional pre-white dwarf lifetime of $\approx$0.3~Gyr \citep{Cummings2018}. Combined, this implies a total age of approximately $\approx$1.3~Gyr. The isochrone age of \thisstarA{} has previously been estimated as $1.6^{+0.8}_{-1.0}$~Gyr \citep{Valenti2005}. Though the uncertainties are relatively large, the consistency of the stellar ages suggests our parameter estimates for \thisstarB{} are reasonable.}

Though \bmaroon{these parameters provide} an adequate model for the observed photometry, the existing data is insufficient to fully characterise \thisstarB{}. In particular, we cannot test our assumption of a hydrogen-dominated atmospheric composition. \bmaroon{Precise} characterisation of \thisstarB{} will require observation of its optical spectrum.

\section{Discussion and conclusions} \label{sec:discussion}

In this work we have presented the serendipitous detection of a white dwarf companion to the star \thisstar{} in JWST/MIRI coronagaphic observations. To our knowledge, this is the first observation of a white dwarf with MIRI using a coronagraphic observing mode; the closest parallel that we know of is the detection of HD~114174~B \citep{Crepp2013, Zhang2023} with NIRCam coronagraphy in JWST commissioning \citep{Girard2022, Kammerer2022}.

With an F-type main sequence primary and a white dwarf secondary, \thisstar{} is an example of a ``Sirius-like system'' as defined by \citet{Holberg2013}. Though in this case \thisstarB{} had already been discovered by \textit{Gaia}, Sirius-like white dwarfs have been quite frequently discovered as a result of coronagraphic imaging observations \citep[e.g.][]{Crepp2013, Crepp2018, Zurlo2013, Hirsch2019, Bonavita2020, Bowler2021}. In the near-IR white dwarfs are as luminous as brown dwarfs, and can easily be confused with them without information from other sources \citep{Mugrauer2005, Crepp2013, Zurlo2013}. In the mid-IR we find that \thisstarB{} is even fainter than the HR~8799 planets \citep{Boccaletti2024}, highlighting the potential for confusion between white dwarfs and planets. We encourage observers to take heed of the possibility that faint co-moving sources detected in MIRI observations could be white dwarfs instead of planets or brown dwarfs. For newly detected objects, care should be taken to distinguish between these possibilities. In particular, as white dwarfs and planets have very different temperatures, broadband photometry at shorter wavelengths can be an efficient way to break this potential degeneracy.

\citet{Holberg2013} found a significant decline in the number of known Sirius-like systems known beyond 20~pc; with a distance of 29~pc, \thisstar{} adds to this under-sampled area of the local Sirius-like population. These systems can act as important benchmarks for aspects of white dwarf physics \citep[e.g.][]{Venner2023}. To improve on our understanding of this system, we advocate for future observation of the optical spectrum of \thisstarB{} to precisely determine the physical parameters of this white dwarf.

\thisstarB{} joins the relatively small list of WDs detected beyond 10~$\mu$m, highlighting the exceptional capabilities of MIRI in this area. Observations of white dwarfs with MIRI are testing the limits of physical models for these stars \citep[e.g.][]{Blouin2024}. Further afield, MIRI provides an unprecedented sensitivity to exoplanets orbiting white dwarfs \citep{Limbach2022} and has been used to discover a small number of candidate planets thus far \citep{Mullally2024, Poulsen2024, Limbach2024}. These observations have previously been conducted with MIRI wide-field imaging and have hence been excluded from observing white dwarfs in close binaries. However, the successful detection of \thisstarB{} in MIRI coronagraphic imaging presented here raises the possibility that the same methods could be applied to planet detection for white dwarfs with IR-bright stellar companions. Of the six white dwarfs within 6~parsecs of the Sun, four are in close binaries ($\lesssim$10") with main sequence stars (Sirius~B; Procyon~B; 40~Eri~B; Gliese~169.1~B). By using the coronagraphs to block out the bright companions, MIRI could conceivably be used to search for white dwarf planets in these and other systems.

\section*{Acknowledgements}

We acknowledge and pay respect to Australia’s Aboriginal and Torres Strait Islander peoples, who are the traditional custodians of the lands, waterways and skies all across Australia. We thank the anonymous referee for their comments that have helped to improve this work. AV would like to acknowledge the ``JWST Master Class 2022'' workshop, hosted at Swinburne University, for bringing the observations of \thisstar{} to their attention. AV is supported by ARC DECRA project DE200101840. LAP acknowledges support from the University of Michigan through the ELT Fellowship Program.

This work is based on observations made with the NASA / ESA / CSA James Webb Space Telescope. Data was obtained from the Mikulski Archive for Space Telescopes at the Space Telescope Science Institute, which is operated by the Association of Universities for Research in Astronomy, Inc., under NASA contract NAS 5-03127 for JWST. These observations are associated with JWST programme \#1037 and \#1194. This work has made use of data from the European Space Agency (ESA) mission {\it Gaia} (\url{https://www.cosmos.esa.int/gaia}), processed by the {\it Gaia} Data Processing and Analysis Consortium (DPAC, \url{https://www.cosmos.esa.int/web/gaia/dpac/consortium}). Funding for the DPAC has been provided by national institutions, in particular the institutions participating in the {\it Gaia} Multilateral Agreement. This publication makes use of data products from the Wide-field Infrared Survey Explorer, which is a joint project of the University of California, Los Angeles, and the Jet Propulsion Laboratory/California Institute of Technology, funded by the National Aeronautics and Space Administration. This research has made use of the SIMBAD database and VizieR catalogue access tool, operated at CDS, Strasbourg, France. This research has made use of NASA's Astrophysics Data System.

\section*{Data Availability}

This paper makes use of observations from the James Webb Space Telescope that are publicly accessible from the Mikulski Archive for Space Telescopes (MAST; \url{https://archive.stsci.edu/missions-and-data/jwst}).




\bibliographystyle{mnras}
\bibliography{bib} 




%
%


\bsp	
\label{lastpage}
\end{document}